\documentclass[12pt]{spieman}  % 12pt font required by SPIE;
\usepackage{amsmath,amsfonts,amssymb}
\usepackage{graphicx}
\usepackage{setspace}
\usepackage{tocloft}
\usepackage{multirow}
\usepackage{float}

\usepackage{upgreek} 
\usepackage[caption=false]{subfig} 

\graphicspath{./Figures/}

\title{Initial Characterization of the First Speedster-EXD550 Event-Driven X-Ray Hybrid Complementary Metal-Oxide Semiconductor Detectors}

\author[a,*]{Joseph M. Colosimo}
\author[a]{Hannah M. Grzybowski}
\author[a]{Evan C. Jennerjahn}
\author[a]{Lukas R. Stone}
\author[a]{Abraham D. Falcone}
\author[a]{Mitchell Wages}
\author[a]{Jacob C. Buffington}
\author[a]{David N. Burrows}
\author[a]{Zachary E. Catlin}
\author[a]{Timothy Emeigh}
\author[a]{Fredric Hancock}
\affil[a]{Pennsylvania State University, Department of Astronomy and Astrophysics, University Park, Pennsylvania, United States, 16802}

\cftpagenumbersoff{figure}
\cftpagenumbersoff{table} 

\newlength{\Oldarrayrulewidth}
\newcommand{\Cline}[2]{%
  \noalign{\global\setlength{\Oldarrayrulewidth}{\arrayrulewidth}}%
  \noalign{\global\setlength{\arrayrulewidth}{#1}}\cline{#2}%
  \noalign{\global\setlength{\arrayrulewidth}{\Oldarrayrulewidth}}}

\begin{document} 
\maketitle

%see here for the original template: https://www.overleaf.com/latex/templates/spie-journal-papers-sample-manuscript-showing-style-and-formatting-specifications/vdydzwfvfvqg 

\begin{abstract}

Future x-ray observatories will require imaging detectors with fast readout speeds that simultaneously achieve or exceed the other high performance parameters of x-ray charge-coupled devices (CCDs) used in many missions over the past three decades. 
Fast readout will reduce the impact of pile-up in missions with large collecting areas while also improving performance in other respects like timing resolution. 
Event-driven readout, in which only pixels with charge from x-ray events are read out, can be used to achieve these faster operating speeds.
Speedster-EXD550 detectors are hybrid complementary metal-oxide semiconductor (CMOS) detectors capable of event-driven readout, developed by Teledyne Imaging Sensors and Penn State University.
We present initial results from measurements of the first of these detectors, demonstrating their capabilities and performance in both full-frame and event-driven readout modes.
These include dark current, read noise, gain variation, and energy resolution measurements from the first two engineering-grade devices. 

\end{abstract}

% Include a list of up to six keywords after the abstract

% Include email contact information for corresponding author
{\noindent \footnotesize\textbf{*}Joseph Colosimo,  \linkable{jcolosimo@psu.edu} }

\begin{spacing}{1}   % use double spacing for rest of manuscript

\section{Introduction}
\label{sect:intro}  % \label{} allows reference to this section

\subsection{X-Ray Hybrid CMOS Detectors}

X-ray Hybrid CMOS detectors (HCDs) are active-pixel sensors composed of separate silicon absorber and readout integrated circuit (ROIC) layers bonded together.
The Penn State High-Energy Astrophysics Detector and Instrumentation Lab has worked with Teledyne Imaging Sensors to develop x-ray HCDs to demonstrate and improve performance for future observatories.
The design of these devices allows for fast and flexible readout architectures compared with the charge-coupled devices (CCDs) used in most currently operating x-ray missions.
The CMOS ROICs also require less power and are more resilient to radiation than CCDs \cite{Bray20}, both critical considerations for space missions.
Furthermore, the hybrid nature of these devices allows for thicker fully-depleted absorbing layers and high quantum efficiency across the soft x-ray band \cite{Bongiorno15, Colosimo22a}.
Measurements on previous versions of x-ray HCDs have demonstrated very rapid readout, low power, and radiation hardness, along with moderate read noise and energy resolution \cite{Prieskorn13, Griffith16, Hull19, Falcone21}, with promising improvements over each iteration.

One of the primary advantages of HCDs and other active-pixel sensors is their rapid readout.
This is particularly important to x-ray detectors operating in the photon-counting regime, as only one photon may land in an individual pixel in each frame in order to be properly registered.
When pile-up of multiple x-rays occurs, the multiple lower-energy photons cannot be distinguished from a single higher energy photon.
This leads to a loss of information about the rate and energies of the observed x-rays.
Faster frame rates decrease the frequency of pile-up events when observing bright sources.
This will be crucial on many next-generation observatories, which will have large collecting areas to increase sensitivity to faint sources, but will still need the capability to produce quality observations of bright sources \cite{Falcone19, Gaskin19}.

Fast readout also provides other advantages, including greater timing resolution and reduced impact from dark current.
Increased timing resolution will be valuable for observations of bright sources exhibiting rapid variability. 
High frame rates also reduce the impact of dark current, as fewer electrons can accumulate in each frame.
The detectors can thus operate at higher temperatures while still achieving performance comparable to that when they are cooled to lower temperatures. 
Furthermore, faster frame rates will also reduce the impact of optical and ultraviolet (UV) backgrounds, reducing the probability of these photons landing coincident to x-rays \cite{Colosimo22b}.
This allows for the use of thinner optical/UV blocking filters, which will provide the instrument with higher quantum efficiency at low energies.

\subsection{Event-Driven Readout} 

Readout speeds can be substantially increased by reading out only those pixels with x-ray events. 
In photon-counting x-ray detectors, the vast majority of pixels in a given frame will contain no charge from x-ray interactions. 
Event-driven readout, in which only pixels with significant charge accumulated in a frame are read out, can thus increase readout speed significantly. 
Penn State University and Teledyne Imaging Sensors have developed this capability in the Speedster-EXD HCDs.

\begin{figure}[bt]
    \centering
    \includegraphics[width=0.8\textwidth]{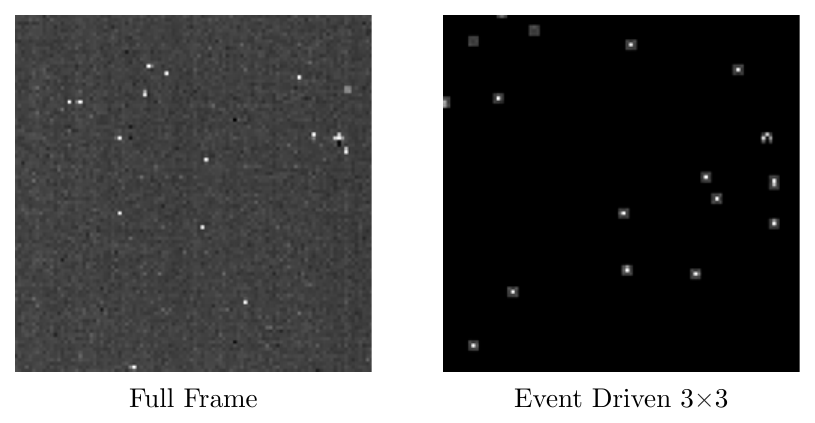}
    \caption{Portions of frames containing Mn K$\upalpha$ and K$\upbeta$ events from an $^{55}$Fe source. The image on the left shows a portion of a frame taken in full-frame readout mode. The image on the right shows another frame taken in the event-driven readout mode, in which only pixels with sufficient signal (and adjacent pixels) are read out. The black regions in this image show pixels which were not triggered for readout.}
    \label{fig:sparse_evts}
\end{figure}

\begin{figure}
    \centering
    \includegraphics[width=0.7\columnwidth]{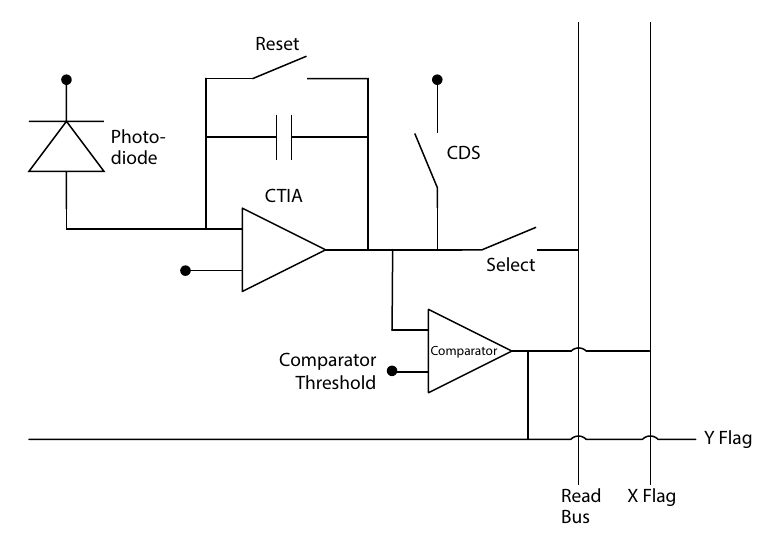}
    \caption{Simplified architecture of a Speedster-EXD pixel. The pixel circuitry is identical in the $64\times64$ and $550\times 550$ arrays.}
    \label{fig:pix_schematic}
\end{figure}

Speedster-EXD HCDs use an in-pixel comparator to flag pixels with significant charge, allowing only those pixels to be read out by the detector.
The comparator threshold is set by the user to specify the amount of charge required to trigger readout. 
These detectors are also capable of reading out the $3\times3$ region of pixels adjacent to all pixels with charge exceeding the comparator threshold. 
The $3\times3$ event-driven readout mode allows for improved energy resolution by enabling the charge to be fully accounted for when the charge cloud overlaps multiple pixels, even when these secondary pixels do not have enough charge to trigger the comparator. 
The detectors can also operate in full-frame or windowed region-of-interest (ROI) readout modes.
Portions of frames captured in both full-frame and $3\times3$ event-driven modes are shown in Fig.~\ref{fig:sparse_evts}.

Fig.~\ref{fig:pix_schematic} shows a simplified schematic of a Speedster-EXD pixel.
Charge is amplified using a capacitive transimpedance amplifier (CTIA), keeping the sense node at a constant potential and avoiding interpixel capacitance observed in previous HCDs using source follower amplifiers \cite{Prieskorn13}.
Pixels also include correlated double sampling (CDS) to reduce the impact of reset noise.
Both reset and CDS are conducted by frame.
These operations contribute to the dead time, during which the detector is insensitive to x-rays.
The comparator is located at the output of the CTIA.
If a pixel's signal exceeds the global comparator threshold, the pixel is flagged for readout.
The comparator threshold voltage is reset each frame.
This operation also contributes to the dead time when the detector is operated in event-driven readout mode.

\begin{figure}[bt]
    \centering
    \includegraphics[width=0.5\textwidth]{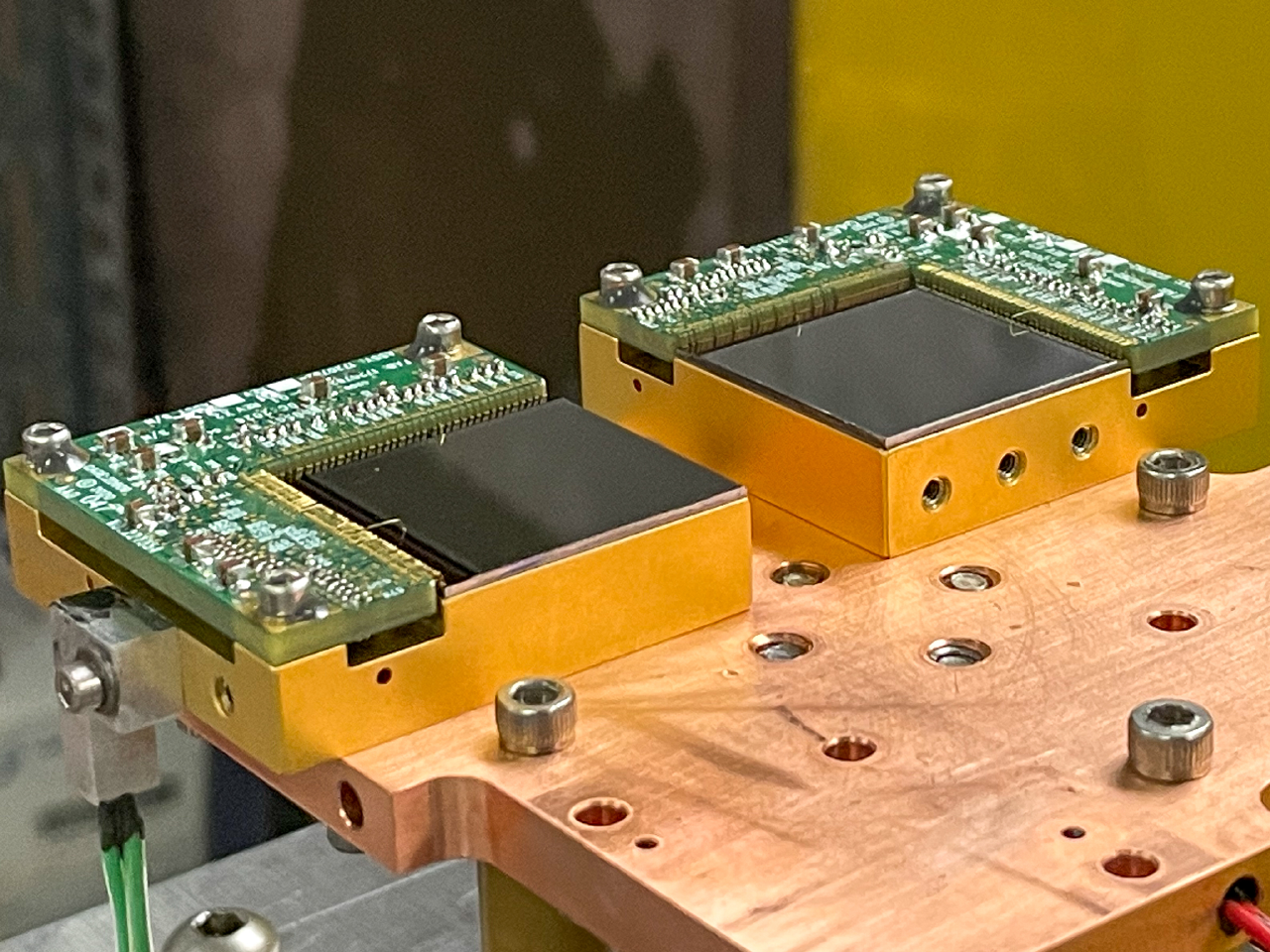}
    \caption{Image of the two Speedster-EXD550 detectors characterized in this work. The detectors are mounted on a copper cold plate.}
    \label{fig:sp550_photo}
\end{figure}

The Speedster-EXD550 is the latest detector developed in partnership between Penn State University and Teledyne Imaging Sensors.
Fig.~\ref{fig:sp550_photo} shows two Speedster-EXD550 detectors mounted in a testing configuration.
These sensors are a scaled-up version of the smaller prototype Speedster-EXD64 detectors \cite{Griffith16}, which have $64 \times 64$ pixel arrays.
The Speedster-EXD550 features a larger pixel array ($550 \times 550$), a 2-side-buttable molybdenum package, and on-chip column-parallel digitization, while maintaining the same unit-cell circuitry as the Speedster-EXD64.
Both versions have a pixel pitch of 40$\upmu$m and a thickness of 100 $\upmu$m.
A substrate voltage of 25 V is applied across the absorber layer to ensure full depletion and minimize the amount of charge spreading across pixels.

\subsection{Speedster-EXD Detectors for the BlackCAT CubeSat}

Rapid readout speed, radiation hardness, and low power draw make these detectors well suited for many applications in space-based x-ray observatories. 
An array of four Speedster-EXD550 detectors will be used in the focal plane array of the BlackCAT CubeSat, whose payload is being designed, assembled, and calibrated at Penn State.
BlackCAT is a 6U NASA CubeSat mission planned to launch in early 2025.
This instrument will use a coded-aperture mask in conjunction with the Speedster detectors to monitor the x-ray sky for gamma-ray bursts and other high-energy transient phenomena.
Fast readout will reduce the impact of dark current, thus allowing for operation at higher temperatures and relaxing the thermal requirements placed upon the limited CubeSat platform \cite{Baker2022}.

\section{Methods}
\label{sec:methods}

We describe initial measurements on two engineering-grade Speedster-EXD550 HCDs, referred to as focal plane modules (FPMs) 23056 and 23057.
We report initial dark current, read noise, gain variation, and energy resolutions for these detectors.
We detail the analysis and results of measurements taken in both full-frame and event-driven readout modes. 
In this section, we describe the test setup and analysis used to characterize the detector performance.

\subsection{Experimental Setup}
\label{sec:setup}

Our measurements are made using a radioactive $^{55}$Fe source, emitting Mn K$\upalpha$ and K$\upbeta$ x-rays at 5.9 and 6.5 keV respectively.  
The detectors can be shielded from this emission using a retractable shutter in order to acquire dark frames. 
The detectors are cooled to 213 K during normal operation to reduce dark current to negligible levels. 
This temperature is achieved by thermally connecting the copper cold mount on which the detectors are mounted (as shown in Fig.~\ref{fig:sp550_photo}) to a liquid nitrogen dewar.
The temperature of the detectors is controlled using heaters potted within the copper plate. 
The temperature setpoint can be changed to allow for characterization of dark current at a range of temperatures.
The measurements took place with the detector assembly in a large vacuum chamber.
Measurements are conducted in vacuum in order to prevent condensation on the cooled detectors and to avoid attenuation of x-rays in air. 

The detectors are operated using a frame grabber and power-supplies, both located outside the chamber.
The detectors are operated at their maximum full-frame readout rate of 152 frames per second for measurements in both full-frame and event-driven readout modes, unless otherwise noted. 
In the test setup used for these measurements, this is the fastest rate that does not drop frames, due to the method in which frames are transferred via the frame grabber. Recent testing with a new detector readout/interface board, developed by the BlackCAT electronics team, has demonstrated operation in a windowed region-of-interest (ROI) mode and event-driven modes at faster rates.
When operated in event-driven readout mode, these detectors are theoretically capable of operating at frame rates in excess of 1 kHz, with the actual rate limited by the number of pixels which must be read out and the desired fraction of dead time in which the detectors are insensitive to x-rays.
With current settings, the dead fraction would be $8\%$ at 1 kHz operation; however, this can be reduced with some impact to read noise performance.

The power dissipation of these detectors is strongly dependent on the settings with which the detector is operated, particularly those related to the operation of the CTIA and the comparator. 
When operated at 213 K with the settings used for the measurements described in this paper, the power draw is 0.42 W for FPM 23056 and 0.45 W for FPM 23057. 
The in-pixel comparator can be disabled with minimal impact to full-frame readout performance. In this state, the power draw can be reduced to 0.21 W for FPM 23056 and 0.22 W for FPM 23057.

\subsection{Image Processing and Event Analysis}

We use a conventional analysis pipeline to remove noise artifacts from images and to identify and characterize x-ray events. 
A set of dark frames is collected in order to remove fixed-pattern noise from images. 
An outlier-resistant truncated mean is calculated for each pixel over these dark frames to create a bias frame used for this purpose.
For data taken in event-driven readout mode, dark frames are acquired at the same effective frame rate in ROI mode. 
For event-driven readout at frame rates faster than the maximum full-frame rate, dark bias frames can be created by combining multiple smaller ROIs acquired separately. 

Frames are observed to have a temporally varying global offset from the computed dark bias mean, adding to the standard read noise.
Images collected in full-frame mode can be corrected for variations in global offset in addition to the standard fixed-pattern noise.
This is done by calculating the offset (the truncated mean of the difference between pixel values and the dark bias frame) and then subtracting this value from the frame. 
This additional step is not done in the analysis of event-driven data, as only pixels adjacent to those with registered charge are read out, biasing this correction. 

X-ray events are identified within the bias-subtracted frames using a conventional algorithm, selecting pixels whose signal is above a set `event' threshold and is greater than that of all adjacent pixels.
The energy of the x-ray event is determined based on the signal of this primary pixel and the signal from charge which spread into surrounding pixels. 
The charge measured in adjacent pixels is added to the energy of an event if it exceeds a `split' threshold.
The number and spatial configuration of these included pixels is used to give the event a quality grade.
We use the event-grading scheme of the \textit{Swift} X-ray Telescope\cite{Burrows05} to classify events.
A bad-pixel mask, created by identifying pixels with anomalous gain or noise, is applied to the event data to mask defective pixels from the analysis.

\section{Results}
\label{sec:results}

\subsection{Gain Variation}
\label{sec:gain_res}

Pixel-to-pixel variation in the gain is a consequence of the detector's design. Every pixel has its own circuitry (signal amplification and readout), and pixel-to-pixel variations of the gain may be present due to component process variations. If left uncorrected, variations in the pixel gain will result in differences in energy measurements of x-rays depending on where the x-ray is absorbed on the detector. Consequently, energy resolution will be degraded unless corrections are made to account for the variations in gain. To correct this effect and properly measure the energy resolution of the detectors presented here, we measured the gain in each pixel of both detectors to fully characterize the gain variation.

The gain variability measurements were made within the experimental setup described in Sec.~\ref{sec:methods}, and x-ray data were collected until $\sim$800 Mn K$\upalpha$ single-pixel events were detected in each pixel. Approximately 20 hours of data collection per detector is required to compile the desired number of single-pixel events per pixel. The captured images were background subtracted and analyzed for single-pixel events using the event-analysis algorithm described above. The resulting single-pixel event values were binned by energy to identify the location of each pixel's Mn K$\upalpha$ and Mn K$\upbeta$ peaks. 

Each pixel's Mn K$\upalpha$ peak was fit with a Gaussian. 
The gain for each pixel was found using the following:
\begin{equation}
\text{Gain} = \dfrac{\mu_{\textrm{Gaussian}}}{5.89\text{ keV}/\omega} \text{ ADU}/e^{-},
\end{equation}
where $\mu_{\textrm{Gaussian}}$ is the mean of the Gaussian fit, in ADU, for the Mn K$\upalpha$ line in a pixel, and $\omega$ is the average energy required for the creation of a single electron-hole pair in silicon which has been accurately measured to be $\omega$ = 3.65 eV/e$^{-}$ \cite{kotov2018pair}. $\omega$ is a property of silicon itself and is slightly dependent on the detector state and operating temperature; however, the temperature dependence is minimal and will have a negligible impact on the results presented.

A gain variation map was then created for each detector. Bad pixels were removed from the gain map and gain variation calculations because these same pixels would be masked out when used in any scientific application. The masked pixels were empty or contained exceptionally high or low gain. These ‘bad-gain’ pixels were removed if they were $\geq 7\sigma$ from the mean of the gain map. The associated gain variation was calculated by dividing the standard deviation of the gain map by the mean.  

Approximately 1.5$\%$ of pixels were masked from the gain variation calculation for FPM 23056, and 5.2$\%$ of pixels were masked for FPM 23057. The larger number of masked pixels in FPM 23057 is due to many columns of dead pixels localized to one side of the detector. The large number of adjacent dead columns of pixels can be explained by manufacturing defects in column readout circuitry. 

\begin{figure}[bt]
    \centering
    \includegraphics[width=0.6\textwidth]{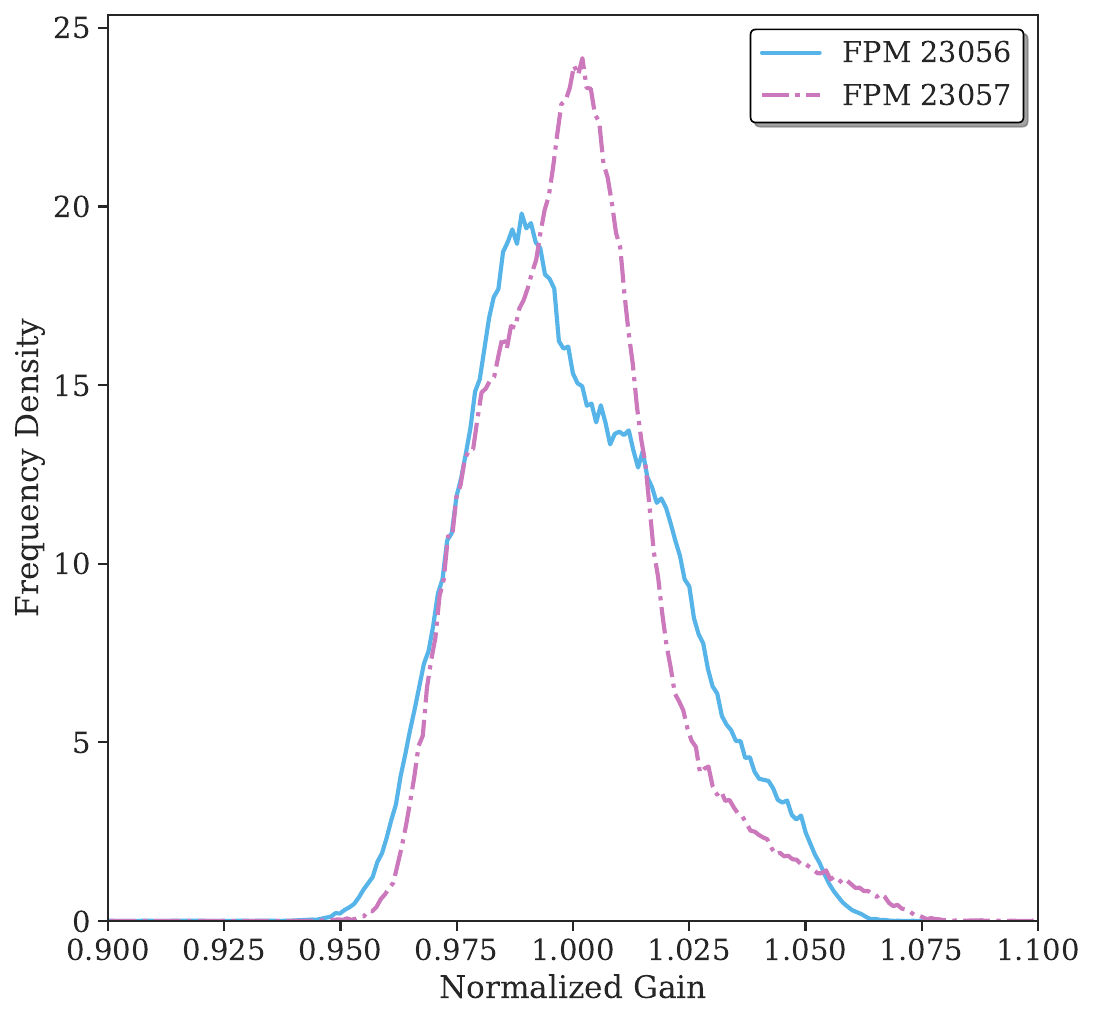}
    \caption{Histograms of the normalized pixel gain values in FPM 23056 (solid blue line) and FPM 23057 (dot-dashed pink line), as measured.}
    \vspace{ 0.75 cm}
    \label{fig:Gain_var}
\end{figure}

The gain variation was measured to be 2.15 $\pm$ 0.09$\%$ for FPM 23056 and 1.95 $\pm$ 0.02$\%$ for FPM 23057. Errors in the gain variation were calculated using the 1$\sigma$ errors on the centroid of the Gaussian fit for each pixel and calculating the mean error of all included pixels.   

Fig.~\ref{fig:Gain_var} shows a histogram of the gain map for FPM 23056 (blue solid line) and FPM 23057 (pink dot-dashed line) normalized by the mean of the respective gain map. 
By examining a spatial map of gain for both devices, we see that some of the gain variation appears to have a spatial dependency.
The exact reason for the localized regions of high or low gain is unknown but may be the result of uneven pressure when the detector was hybridized. 
Moreover, pixels around non-uniform regions, i.e., dead pixel groups or columns, appear to suffer from high deviations from the mean.
Overall, this gain variation is modest and is easily measured, thus enabling a pixel-by-pixel gain correction to be applied to the measured signal in each pixel.

\subsection{Dark Current}
\label{sec:dark_current}

\subsubsection{Methodology}
In order to characterize the dark current of these detectors, we measured the dark current at 10 degree increments from 203 K to 263 K. 
At each temperature, dark frames were acquired at several different exposure times. 
After applying a bad-pixel mask, we applied a linear fit to each pixel's signal as a function of the exposure time. The slope of that line provides the dark current for that pixel at that temperature. 
The dark current of the detector is quoted as the median of the pixel dark currents. 
The quoted errors give the median absolute deviation (scaled for equivalence to standard deviation) of the measured dark current for each pixel. 

We acquired $^{55}$Fe data at each temperature in order to properly correct the dark current measurements for changes in the overall detector gain as a function of temperature. 
These data were used to identify the location of the centroid of the Mn K$\upalpha$ peak in order to measure and account for changes in gain as a function of temperature in our dark current measurements.
We do not correct for pixel-to-pixel gain variation in these dark current measurements due to the substantial time required to acquire enough single-pixel Mn K$\upalpha$ events to produce a proper pixel gain map at each temperature.
The overall detector gain is sufficient to scale the measured dark current, and gain variation is small enough to only marginally increase the scatter observed in pixel dark current.

\subsubsection{Measurements and Model}

\begin{table}[bt]
\caption{Dark current measurements for FPM 23057 and FPM 23056.} 
\label{tab:DC values}
\begin{center}       
\begin{tabular}{lll} %% this creates two columns
%% |l|l| to left justify each column entry
%% |c|c| to center each column entry
%% use of \rule[]{}{} below opens up each row
\hline
\hline
\rule[-1ex]{0pt}{3.5ex}  Temperature (K) & FPM 23056 (e$^-$/pix/s) & FPM 23057 (e$^-$/pix/s) \\
\hline\hline
\rule[-1ex]{0pt}{3.5ex}  203 & $37.5 \pm 4.0$ & $48.5 \pm 4.8$ \\
\rule[-1ex]{0pt}{3.5ex}  213 & $239 \pm 16$ & $279 \pm 22$ \\
\rule[-1ex]{0pt}{3.5ex}  223 & $1256 \pm 80$ & $1390 \pm 100$ \\
\rule[-1ex]{0pt}{3.5ex}  233 & $5500 \pm 340$ & $6100 \pm 420$ \\
\rule[-1ex]{0pt}{3.5ex}  243 & $23400 \pm 1400$ & $22500 \pm 1500$ \\
\rule[-1ex]{0pt}{3.5ex}  253 & $87300 \pm 5300$ & $87800 \pm 5700$ \\
\rule[-1ex]{0pt}{3.5ex}  263 & $284000 \pm 18000$ & $296000 \pm 18000$ \\
\hline 
\end{tabular}
\end{center}
\end{table} 

The dark currents found at each temperature for both detectors were recorded along with their error in Table \ref{tab:DC values}.
The dark current is expected to increase with temperature according to the following equation\cite{janesick_2001}:
\begin{equation}
\label{eqn:DC}
\mathrm{Dark\ Current(e^-/pix/s)}=2.5\times10^{15} P_{s} \, D_\mathrm{FM} \, T^{1.5} \, e^{-E_g /2kT},
\end{equation}
where $P_s$ is the pixel area in cm$^2$, $D_{FM}$ is the dark current figure of merit at 300 K in nA/cm$^2$, $T$ is temperature in K, $k$ is Boltzmann's constant in eV/K, and $E_g$ is the silicon band-gap energy in eV, which is a function of temperature per the following equation \cite{janesick_2001}:
\begin{equation}
\label{eqn:Eg}
\mathrm{E_g{(eV)}}=1.557-\frac{7.021\times10^{-4}\,T^2}{1108+T}.
\end{equation}

\begin{figure}[t]
    \centering
    \includegraphics[width=0.95\textwidth]{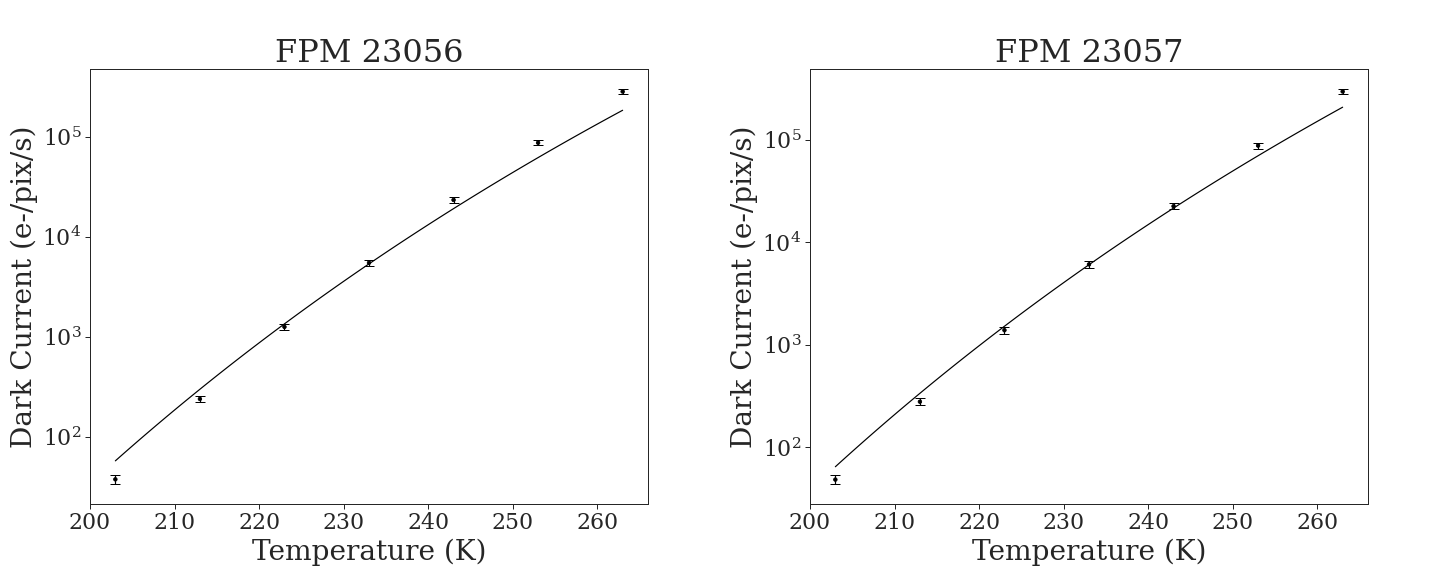}
    \caption{Dark current as a function of temperature for FPM 23056 and FPM 23057 along with model fits.}
    \label{fig:Dark Current Vs. Temp Plots}
\end{figure}

Eq.~\ref{eqn:DC} accounts for both surface and depletion dark current, which are expected to be the dominant sources of dark current in these detectors. Both detectors approximately follow the expected behavior from 203 K to 263 K with some departures from the model at both extremes, so we can use the model from Eq.~\ref{eqn:DC} to fit these data. Every variable in Eq.~\ref{eqn:DC} is known with the exception of the figure of merit at 300 K, which is determined by the fits shown in Fig.~\ref{fig:Dark Current Vs. Temp Plots}. 
This results in a figure of merit of 58.8 nA/cm$^2$ for FPM 23056 and 66.1 nA/cm$^2$ for FPM 23057.

\subsubsection{Expected Impact of Dark Current: A Sample Case}
The BlackCAT CubeSat will utilize passive cooling, with expected focal plane operating temperatures between 213 K and 233 K once in orbit \cite{Baker2022}.
It is particularly important to characterize the level and impact of dark current in the Speedster-EXD550 detectors when operating within this temperature range. 
The shot-noise contribution from dark current is the square root of the expected number of dark-current electrons accumulated in a pixel in a single frame. 
We find that at our fast frame rates (maximum of 152 Hz for full-frame observations), the dark current will not be the dominant source of noise. At 233 K, the expected noise from dark current is 6.0 e$^-$/pixel for FPM 23056 and 6.3 e$^-$/pixel for FPM 23057 at frame rates of 152 Hz.
These values are substantially lower than the read noise measured on these devices, contributing only marginally to the overall detector noise when cooled to 233 K. 
The impact of dark current will be reduced even further in event-driven readout mode because the frame rate can be substantially increased.

\subsection{Read Noise}
\label{sec:read_noise}
The read noise is characterized by measuring the variation of pixel signals acquired in dark frames.
We quote the noise as the root mean square (RMS) of the signal of an individual pixel over a large number of frames. 
The frames are corrected for offset fixed-pattern noise by calculating and subtracting bias frames prior to our noise analysis. 
Data taken in full-frame mode is also corrected for global offset variations, while data taken in event-driven mode is not, resulting in a higher effective read noise for event-driven data than for full-frame data. 
We characterize the effective noise for data acquired in both full-frame and event-driven readout modes.
Measurements were taken with the detectors cooled to 213 K, resulting in little impact of dark current on the measured noise ($\sim1.3$ e$^-$ shot noise from dark current).

The median magnitude of the global offset variation is measured to be approximately 5 e$^-$ in both devices. 
This contributes to the higher effective read noise measured in event-driven readout mode.
The source of this offset noise is currently not well understood.
Future efforts will be aimed at characterizing the dependence of the offset variation magnitude on detector parameters in order to reduce the impact and better identify the source. 

\begin{figure}[tb]
    \centering
    \includegraphics[width=0.95\textwidth]{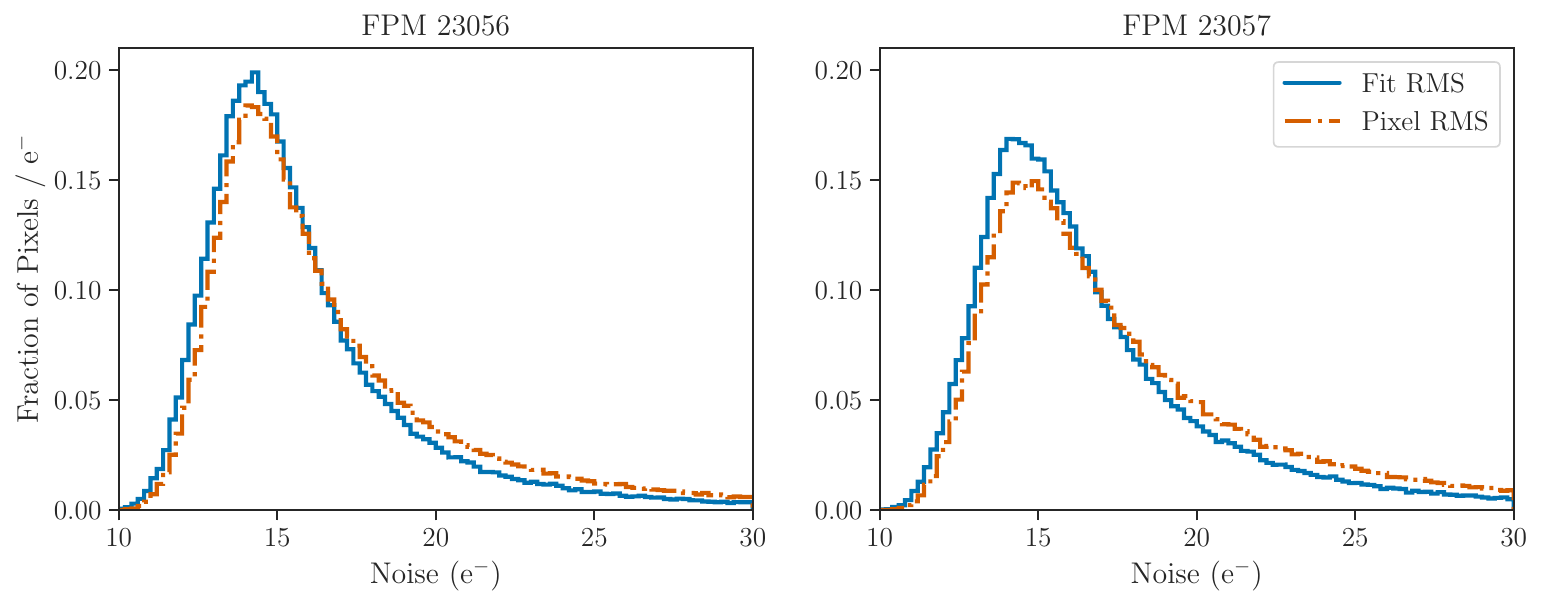}
    \caption{Distribution of noise values for the pixels in both detectors. These noise distributions are from full-frame images corrected for fixed pattern noise and global frame offset variations.}
    \label{fig:pix_noise}
\end{figure}

We measure the noise for each pixel using two different methods. 
We measure the pixel RMS, defined as the RMS of the signal of an individual pixel over a series of 1000 frames.
We also measure the fit RMS, defined as the width of a Gaussian fit to the histogram of a pixel's values taken over the same series of frames.
The pixel RMS gives the actual observed noise of the pixel, while the fit RMS gives the contribution of the Gaussian component. 
Fig.~\ref{fig:pix_noise} shows the distribution of measured pixel noise values for full-frame operation.
Table~\ref{tab:noise} gives the median noise measured for both detectors when operating in both full-frame and event-driven readout modes.

These detectors contain a sizable fraction of pixels exhibiting random telegraph noise (RTN) \cite{Bogaerts02, Janesick06}, with $4.6\%$ of pixels in FPM 23056 and $2.4\%$ in FPM 23057 demonstrating clear signatures of RTN.
These pixels are identified through the presence of multiple distinct noise peaks appearing in histograms of a pixel's values over a large number of frames.
This effect can decrease the energy resolution of a pixel and also can create spurious events if the magnitude of the RTN is large enough. 
Pixels with high-frequency or high-magnitude RTN can be removed from analysis to reduce these effects in order to improve detector performance, particularly at lower energies.
Pixels with RTN can also be deselected from readout in event-driven mode; however, this feature is not currently functional in a large number of columns on both detectors. 
It is also possible that low-magnitude RTN, difficult to distinguish from ordinary read noise, is present in significantly more pixels and is responsible for the large fraction of pixels with read noise measured well above the median value. 

\begin{table}[tb]
\caption{Read noise measurements for both devices operating in full-frame and event-driven readout modes. Correction for variation in global pixel offset is applied in full-frame data but not in event-driven data. Median pixel values are quoted along with 68th-percentile ranges.} 
\label{tab:noise}
\begin{center}       
\begin{tabular}{lll}
\hline
\hline
\rule[-1ex]{0pt}{3.5ex}   & Pixel RMS (e$^-$) & Fit RMS (e$^-$)  \\
\hline\hline
\rule[-1ex]{0pt}{3.5ex}  \textbf{FPM 23056}  \\
\rule[-1ex]{0pt}{3.5ex}  Event Driven & $17.5^{+5.0}_{-1.9}$ & $17.1^{+3.6}_{-1.8}$ \\
\rule[-1ex]{0pt}{3.5ex}  Full Frame & $15.6^{+5.5}_{-2.1}$ & $15.1^{+3.8}_{-1.9}$ \\
\hline
\rule[-1ex]{0pt}{3.5ex}  \textbf{FPM 23057}  \\
\rule[-1ex]{0pt}{3.5ex}  Event Driven & $18.6^{+6.8}_{-2.5}$ & $18.0^{+4.6}_{-2.1}$ \\
\rule[-1ex]{0pt}{3.5ex}  Full Frame & $16.7^{+7.4}_{-2.8}$ & $15.8^{+4.9}_{-2.2}$ \\
\hline 
\end{tabular}
\end{center}
\end{table}

\subsection{Energy Resolution}
\label{sec:energy_res}

To determine the energy resolution of these detectors, we measure the energy width of the Mn K$\upalpha$ peak in data collected using an $^{55}$Fe source. 
Data with $^{55}$Fe x-ray events were acquired in both event-driven and full-frame readout modes, with the detector operating at a temperature of 213 K. 
We subtracted the dark bias frames for all data and the global frame offsets for full-frame data.
We performed event identification using a 10$\sigma$ event threshold (based on the median pixel-RMS noise).
Events from pixels flagged in the bad-pixel mask were discarded from this analysis. 

The energy resolution depends on multiple factors in both detector operation and analysis.
The energy resolution will be different in full-frame and event-driven readout due to the higher effective read noise in event-driven operation. 
The energy resolution is also dependent on the level of the split threshold and the event grades included in the analysis. 
We perform our analysis for split thresholds of 1, 2, and $3\sigma$ and for different selections of event grades.
In the \textit{Swift} XRT grading scheme \cite{Burrows05}, grade-0 events only include a single primary pixel.
These events are generally the highest quality because only noise from a single pixel impacts the measured energy.
Events with grades 1--4 are a 2-pixel split events, and those with grades 5--12 are 3- or 4-pixel events. 
Events with grades $>12$ are not generally used in analysis as these do not correspond with shapes expected from an x-ray interaction.
Once the proper event-selection cuts are applied, we bin events by energy to create spectral energy distributions.

\begin{table}[btp]
\caption{Energy resolution measurements for FPM 23056, giving the FWHM of the Mn K$\upalpha$ peak in eV. Three factors are investigated to determine their impact on the energy resolution: readout mode, included grades (defined by the \textit{Swift} XRT grading scheme), and split threshold (given as a multiple of the pixel-RMS read noise).}
\vspace{0.2 cm}
\label{tab:energyres56}
\resizebox{\columnwidth}{!}{%
\begin{tabular}{cclccccccccccc}
                            & Split Threshold &  & \multicolumn{3}{c}{$1\sigma$} &  & \multicolumn{3}{c}{$2\sigma$} &  & \multicolumn{3}{c}{$3\sigma$} \\  \Cline{2pt}{2-14} 
                            &                     &  &             &  &              &  &             &  &              &  &             &  &              \\
                            & Grades               &  & $\Delta E$ (eV)  &  & \% of Events &  & $\Delta E$ (eV) &  & \% of Events &  & $\Delta E$ (eV) &  & \% of Events \\
                            &                &  & (FWHM)  &  &  &  & (FWHM) &  &  &  & (FWHM) &  &  \\ \Cline{2pt}{2-2} \cline{4-4} \cline{6-6} \cline{8-8} \cline{10-10} \cline{12-12} \cline{14-14} 
                            \\
\multirow{3}{*}{Event Driven}     & 0                   &  & 238         &  & 11\%         &  & 257         &  & 26\%         &  & 279         &  & 33\%         \\
                            & 0-4                  &  & 300         &  & 35\%         &  & 306         &  & 66\%         &  & 316         &  & 76\%         \\
                            & 0-12                 &  & 343         &  & 50\%         &  & 330         &  & 81\%         &  & 332         &  & 88\%         \\
\\
\cline{1-14}
\\

\multirow{3}{*}{Full Frame} & 0                   &  & 223         &  & 8\%         &  & 239         &  & 22\%         &  & 261         &  & 29\%         \\
                            & 0-4                  &  & 274         &  & 28\%         &  & 278         &  & 58\%         &  & 289         &  & 67\%         \\
                            & 0-12                 &  & 308         &  & 43\%         &  & 297         &  & 73\%         &  & 302         &  & 79\%        
\end{tabular}%
}
\vspace{0.25 cm}
\end{table}

%FPM23057

\begin{table}[btp]
\caption{Energy resolution measurements for FPM 23057, similar to those provided in Table \ref{tab:energyres56} for FPM 23056.}
\vspace{0.2 cm}
\label{tab:energyres57}
\resizebox{\columnwidth}{!}{%
\begin{tabular}{cclccccccccccc}
                            & Split Threshold &  & \multicolumn{3}{c}{$1\sigma$} &  & \multicolumn{3}{c}{$2\sigma$} &  & \multicolumn{3}{c}{$3\sigma$} \\  \Cline{2pt}{2-14} 
                            &                     &  &             &  &              &  &             &  &              &  &             &  &              \\
                            & Grades               &  & $\Delta E$  &  & \% of Events &  & $\Delta E$  &  & \% of Events &  & $\Delta E$  &  & \% of Events \\
                            &                &  & (FWHM)  &  &  &  & (FWHM) &  &  &  & (FWHM) &  &  \\ \Cline{2pt}{2-2} \cline{4-4} \cline{6-6} \cline{8-8} \cline{10-10} \cline{12-12} \cline{14-14} 
                            \\
\multirow{3}{*}{Event Driven}     & 0                   &  & 253         &  & 13\%         &  & 277         &  & 27\%         &  & 301         &  & 34\%         \\

                            & 0-4                  &  & 323         &  & 40\%         &  & 328         &  & 68\%         &  & 337         &  & 75\%         \\
                            
                            & 0-12
                 &  & 365         &  & 55\%         &  & 350         &  & 81\%         &  & 351         &  & 85\%         \\
\\
\cline{1-14}
\\

\multirow{3}{*}{Full Frame} & 0                   &  & 237         &  & 10\%         &  & 260         &  & 23\%         &  & 284         &  & 29\%         \\
                            & 0-4                  &  & 295         &  & 33\%         &  & 300         &  & 59\%         &  & 311         &  & 66\%         \\
                            & 0-12                 &  & 330         &  & 49\%         &  & 319         &  & 72\%         &  & 325         &  & 76\%        
\end{tabular}%
}
\vspace{0.25 cm}
\end{table}

\begin{figure*}[tb]
    \centering
    \includegraphics[width=0.7\columnwidth]{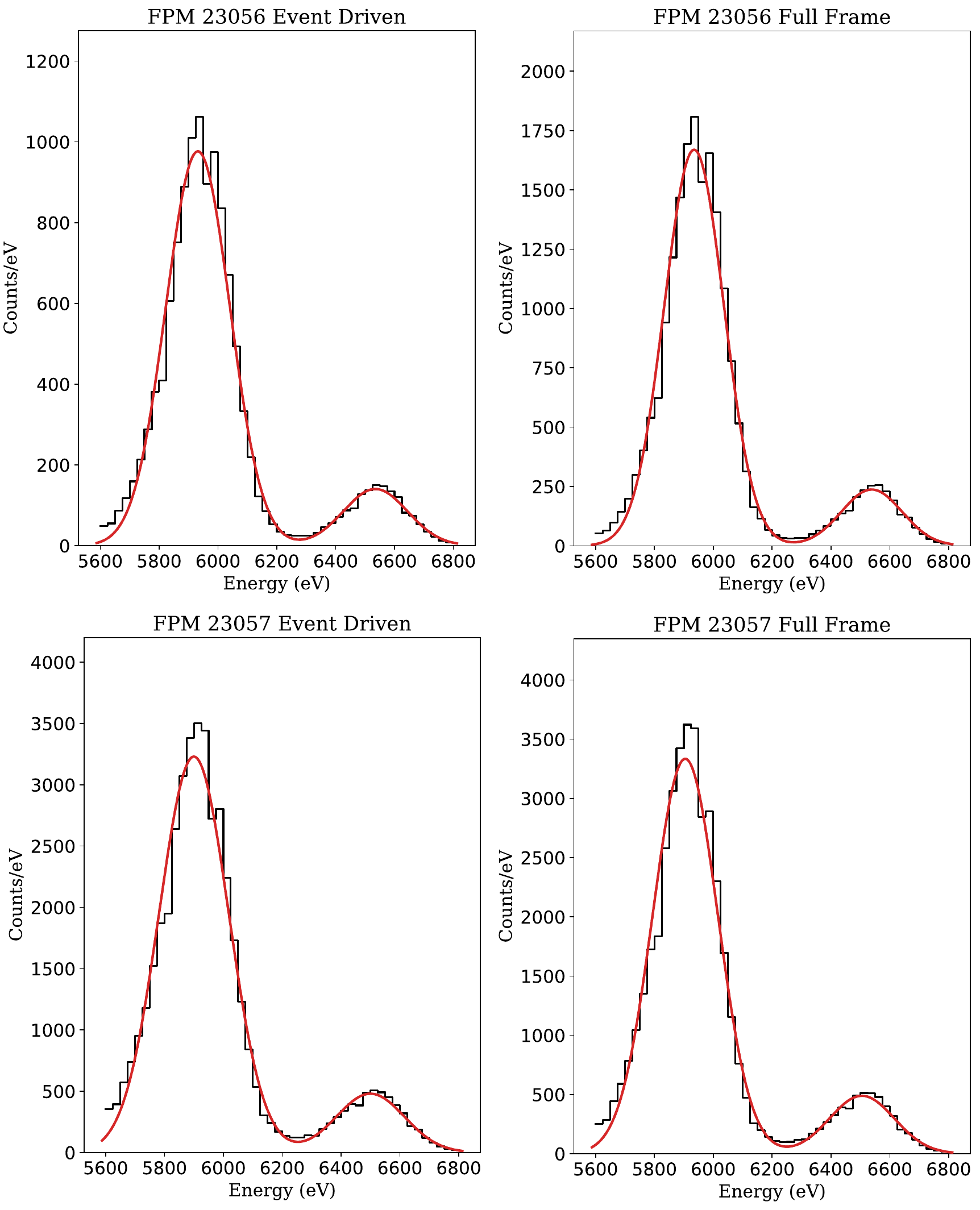}
    \caption{Gaussian fits of Mn K$\upalpha$ and Mn K$\upbeta$ peaks for FPM 23056 and FPM 23057 in both event-driven and full-frame readout modes with grade-0 (single-pixel) events and a split threshold of 2$\sigma$. The FWHM of the Mn K$\upalpha$ Gaussian component of the model is used in reporting the energy resolution.}
    \label{fig:fits}
\end{figure*}

The resulting spectra are fit using two Gaussians, one to model the Mn K$\upalpha$ peak and the other to model the K$\upbeta$ peak.
We quote the full width at half maximum (FWHM) of the Mn K$\upalpha$ peak derived from this fit as the energy resolution.
Example fits from the grade-0, $2\sigma$ split-threshold analysis are shown in Fig.~\ref{fig:fits}.
Full results are displayed in Tables \ref{tab:energyres56} \& \ref{tab:energyres57}.

With a low split threshold, a wider range of acceptable pixel values can be included in the event. 
This can result in the inclusion of pixels adjacent to the primary pixel due to noise rather than actual charge generated by an x-ray. 
A low split threshold will thus produce worse energy resolution for multi-pixel events. 
A high split threshold, on the other hand, can result in signal from pixels into which an insufficient amount of charge has spread to be excluded from the total energy of the event.
This results in an undercounting of the total charge in an event and a degradation of the energy resolution, especially in the case of low-energy x-rays. 
These effects can be observed in Tables \ref{tab:energyres56} \& \ref{tab:energyres57}.
We also note that the higher effective read noise encountered in event-driven readout results in slightly worse energy resolution.

The Speedster-EXD550 devices investigated here show similar performance to the Speedster-EXD64 devices characterized in Ref.~\citenum{Griffith16}, with somewhat higher read noise and lower energy resolution.
The Speedster-EXD64 devices were operated at a lower temperature (150 K) and higher frame rate (1 kHz) for their characterization measurements.
Read noise measurements of 13.0 e$^-$ and 11.2 e$^-$ are reported for the two Speedster-EXD64 devices.
Dark current contributes minimally to the noise in the measurements on the Speedster-EXD550 or Speedster-EXD64 devices; however, other temperature-dependent sources of noise may contribute to the higher noise in the Speedster-EXD550 measurements at 213~K presented in this paper.
For the Speedster-EXD64 detectors, the Mn K$\upalpha$ FWHM was measured to be 206 eV and 236 eV in full-frame operation and 236 eV and 247 eV event-driven operation for single-pixel events (with a $3\sigma$ split threshold).
A full pixel-to-pixel gain correction was not applied to these detectors to obtain these measurements, although the smaller array size and method of fitting these peaks reduces the impact of gain variation on the energy resolution of the Speedster-EXD64 detectors.

\section{Conclusion}
\label{sec:conclusion}

We have presented measurements of two engineering-grade Speedster-EXD550 HCDs, demonstrating their capabilities when operated in both full-frame and event-driven readout modes. 
Fast operation allows these detectors to function at higher temperatures than is feasible for most other HCDs, due to the reduced impact of dark current at these frame rates.
We measured a median read noise of 15.6 and 16.7 e$^-$ (RMS) when operated in full-frame readout and a median effective read noise of 17.5 and 18.7 e$^-$ (RMS) when operated in event-driven readout for each of the respective detectors. 
These detectors have significantly higher read noise than other HCDs, such as the small-pixel detectors, which have measured read noise values as low as 6.3 e$^-$ (RMS) \cite{Hull19}. 
While future x-ray missions will likely require reduced read noise and improved energy resolution, these detectors meet the energy resolution requirements for the BlackCAT CubeSat, which requires fast frame rates but has only modest read noise requirements.

Through continued optimization and characterization at colder operating temperatures, we will attempt to demonstrate lower-noise operation with these devices. 
An ongoing effort is also aimed at designing Speedster-EXD HCDs with lower sense-node capacitance in order to achieve lower read noise while still maintaining the capability for event-driven readout.
Continued testing of these devices with new readout electronics for BlackCAT will also demonstrate event-driven performance at faster frame rates.
Efforts are also underway to characterize the performance of six other Speedster-EXD550 devices from Teledyne Imaging Sensors in preparation for the selection and calibration of detectors for the BlackCAT focal plane. 

% \disclosures 
\subsection*{Disclosures}
%Conflicts of interest should be declared under a "Disclosures" header. If the authors have no relevant financial interests in the manuscript and no other potential conflicts of interest to disclose, a statement to this effect should also be included in the manuscript.
No potential conflicts of interest have been identified by the authors. 

\subsection* {Code, Data, and Materials Availability} 

All data in support of the findings of this paper are available either within the article itself or as supplementary material, which may be provided upon request.

\subsection* {Acknowledgments}
This work was supported by a NASA Space Technology Graduate Research Opportunity (grant 80NSSC20K1210) as well as NASA grants 80NSSC18K0147, NNX14AH68G, and 80NSSC21K1125. 
We would also like to thank Vincent Douence at Teledyne Imaging Sensors for his guidance in operating these detectors and optimizing their performance.

%%%%% References %%%%%

\bibliography{report}   % bibliography data in report.bib
\bibliographystyle{spiejour}   % makes bibtex use spiejour.bst

\end{spacing}
\end{document}